# ICT and RFID in Education: Some Practical Aspects in Campus Life

Cristina TURCU, Cornel TURCU, Valentin POPA, Vasile GAITAN

University of Suceava, 13 University Street, 720229-Suceava, Romania
{cristina, cturcu, valentin, gaitan}@eed.usv.ro

**Abstract**

The paper summarizes our preliminary findings regarding the development and implementation of a newly proposed system based on ICT and RFID (Radio Frequency Identification) technologies for campus access and facility usage. It is generally acknowledged that any educational environment is highly dependent upon a wide range of resources or variables such as teaching staff, research and study areas, meeting and accommodation facilities, library services, restaurant and leisure facilities, etc. The system we have devised using ICT and RFID technologies supports not only authentic transactions among all university departments, but also interconnects all levels of academic life and activity. Thus, the utility of the system ranges from access control (student/ staff/ visitor identification), attendance tracking, library check-out services and voting to grade book consulting, inventory, cashless vending, parking, laundry and copying services. Physically, the system consists of several RFID gates/readers, a data server and some network stations, all of them requiring specific structuring and integration solutions. The system is quite different from already existing ones in that it proposes an innovative access solution. Thus, the search of the ID card holder in a database has been replaced by local processing. Since one and the same card is employed to perform a variety of operations, the system has immediate and numerous utilizations.

## 1. Introduction

The growth in the number of students and the need for high-quality services in Campus have determined us to look for new methods allowing us to rapidly process any service request without hiring additional staff. One way towards such an improvement is to facilitate the quick transfer of information from the supplier of services to the client. The existence of various categories of customers or clients, and hence of different fees for the same service, inevitably requires the adoption of a method ensuring the easy identification of clients or customers and their proper payments in the correct amounts.

The most adequate technology to fulfill this task is RFID [1][2], which is a wireless technology that uses radio communication to identify objects with a unique electrical identity. Widely employed in business and industry, this technology has been particularly favoured by retailers, suppliers and transport providers. The huge advantage of an RFID system resides in that it allows information to be added to or modified in the card or tag memory even at a gate reader [3]. Moreover, administrators can immediately ascribe new access rights and even update data input, e.g. the addition of new accounts for new campus services or the inclusion of emergency medical information.

## 2. The presentation of the system

In what follows, we shall attempt a brief presentation of the incorporated generalized system based on RFID, which was specifically designed to control campus access. Its major characteristic is that it can simultaneously control the access in different campus areas (through gates of access) and even adopt selective control at different times. All users' relevant data may be easily defined and stored. Moreover, users can devise their own structural models of data and then use them as information storage templates on transponders.

### 2.1. Architecture

In Figure 1 we present the architecture of the developed system.

The elements composing the integrated system are:

- a PC which coordinates the whole system
- fixed readers, which can be:
  - PC-connected readers allowing users to perform different tasks;
  - fixed readers made up of embedded devices (terminals) to which readers are connected. As all terminals are connected to a network, every is able to communicate with the PC (through RS232) and with every other terminal in the network (through RS485). When more terminals are needed, a master terminal may be used for up to 30 terminals; a wired or radio communication may link the master terminal to the PC. A special command for local processing is required if terminals are supposed to work independently from the computer.
- mobile readers consisting of a reader connected to a PDA.

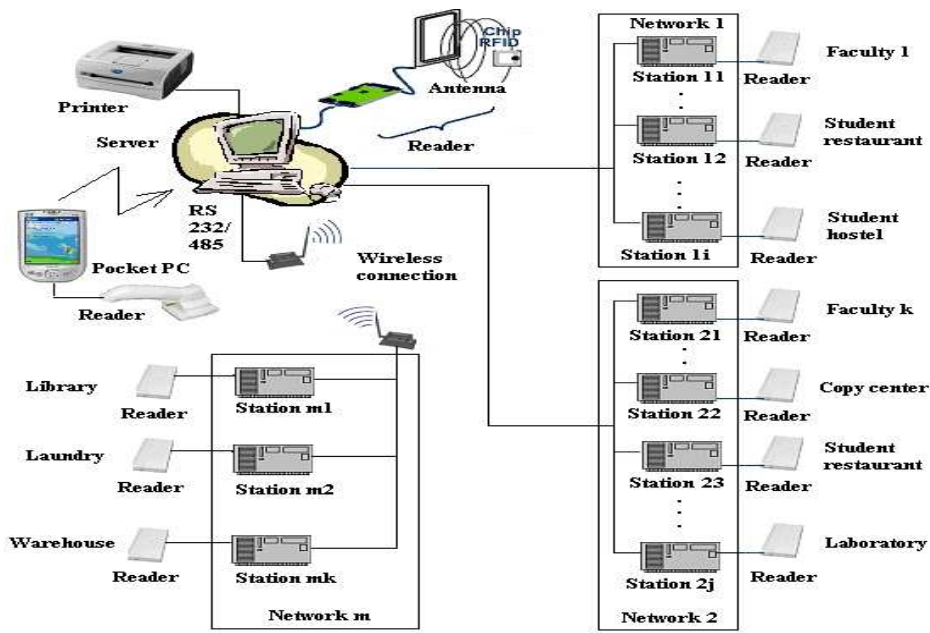

**Figure 1. The RFID based campus system architecture**

The system runs on passive transponders that meet the ISO standard 15693 for frequencies of 13.56 MHz. The transponder serial number contributes to the unique identification of the transponder, which thus acquires a unique identification number. The reading range varies from 9 cm to 40 cm, depending on the reader, the aerial and the transponder. The system affords the simultaneous reading of various transponders; however, the number of simultaneously read transponders may vary among readers   Let us consider the functioning of the system in more detail. Connected to the PC, the first type of fixed reader allows the initial registration of transponders, the reading of those of interest at some point, the visualization of data, as well as the modification of input data. The PC holds the whole transponder database, which is constantly updated with information coming from both fixed and mobile readers. Moreover, the PC is the informative repository of any modification undergone by transponders e.g. transponders which were read/written by using a fixed or perhaps a portable reader.

The second type of fixed reader is the one made up of embedded devices (terminals) to which readers are connected. A terminal is a complex system equipped with a microprocessor, memory, I/O ports, etc. which perform a series of tasks. In order to ensure the general character of the system (i.e. the processing of a large number of transponders), we have decided to allow the information to be stored on the transponders and not in some terminal database. When a transponder enters the action range of the reader (or when a person carrying a transponder approaches the reader), the terminal commands the reading of the transponder and processes the information

(hour, gate list, etc.) allowing or restricting the access through that gate. Depending on the current computer settings, the terminal can store the information on the transponder, can write new information on the transponder or can update the existing information on a precise field on the transponder (e.g., it can change the schedule that corresponds to a transponder). The read/stored data on the terminals are conveyed to the application running on the PC, which will update the database. The events recorded at the terminals are also stored in the terminal memory and they are kept there until the PC application specifically requires them. The terminal receives the commands given by the PC, confirms the receiving of the command or conveys an error message if the command is unknown (e.g. when an error occurs in the transmission of a certain command). If the command is correct, the terminal fulfills it and the response containing the data that resulted after fulfilling the command is conveyed.

Another type of reader is the mobile reader; it consists of a PDA to which a reader has been connected. Mobile readers represent a more flexible approach since the reading and writing of transponder are not spatially or temporally conditioned.

**2.2. System capabilities**

The following operations are easily carried out within the incorporated developed system:
- ✓ the real-time monitoring of doors;
- ✓ the acknowledgement in real time of all occurring events;
- ✓ the generation of reports concerning the activity carried out in a specific area, at a certain time or for a certain person and their conveyance in different formats;
- ✓ the sorting out of events depending on certain criteria and the display of these criteria on the monitor or their printing;
- ✓ the locking/unlocking of a transponder;
- ✓ restricting access;
- ✓ brief unlocking, temporary unlocking or tag-driven unlocking (or certain category-driven unlocking);
- ✓ database history and backup;
- ✓ automatic detection of the terminals in the system;
- ✓ ensuring the security of the system through the control of the access to the system and the coding of the data existent in the database (PC, PDA).
- ✓ the management of the users registered in the system (user visualization, user addition or deletion, modifying the user's profile);
- ✓ ensuring the management of the database which stores the information concerning the transponders, terminal commands, events coming from the terminals, users, system-based events;
- ✓ customizing the incorporated system according to needs, including the customizing of:
    1) the applications running on the PC, the setting of the folder containing the information on the database, the communication rate of the terminals, the display colour;
    2) the terminals, e.g. the time setting of the electromagnetic shutdown system, of the automatic closing of a door left open, of the communication rate, the terminal address, the password, etc.
    3) the PDA applications, after the folder where the database information is stored.
- ✓ special warnings in keeping with the settings and the visualization of the alarms produced in the system;
- ✓ the creation of the layouts to be used for the storage of information on transponders;
- ✓ the management of the database with information regarding the transponders, terminals, etc.
- ✓ ensuring communication with the system clients, e.g. on-line information publishing
- ✓ the storage of the whole information processed by the system and its management (search, find and view).

The transponder may carry a variety of information. Here are several categories: Personal ID number, Card expiry date, Data access block, Card issue number, Holder type (personnel/student), Meal plan, Restaurant account, Service account 1, Medical record 1, Tax payment record.

**2.3. Security features**

The security of the system may be approached from several directions. First, it the transponders that make it secure. Even if the unique serial numbers represent one important element of security, further security measures have been seriously considered. In this respect we have agreed on the identification of registered transponders through our own developed software in order to reduce the risk of the system's acceptance of transponders registered through alien methods. The information stored on the transponder is **exclusively dependent** upon our own developed system and its correspondent layout of data.

Another security method consists in implementing different access routes within the system, using an account and a password.

Moreover, the database security within the system has been enhanced. Encoding algorithms have been employed in encrypting our databases. In addition, all databases are protected with passwords; hence the information entered cannot be altered, for instance, by a management database system.

## 3. The advantages of the system

Allowing further developments and extensions, the present system makes our campus a highly secure environment. At the same time, it can be easily adjusted to meet a variety of campus needs. A highly customizable product, our system presents itself as an extremely useful and flexible solution in managing various aspects of campus life and activity. It has been designed in such a way as to be able to operate at its full capacity even if transponders might undergo profound typological transformations (e.g. if a spectacular evolution takes place on the RFID market, through the introduction of new transponders which are very cheap and have a great storage capacity).

The system ensures an appropriate functioning speed and performance rate for the processing of data coming from thousands of transponders in thousands of different places. In addition, local data processing facilitates the access through a gate and thus the need to interrogate a database server of a centralized system has been eliminated. Of notable interest is the ease of ascribing access rights either from the PC or from any terminal, following a PC command.

Our developed system is not only applicable to managing campus access. Other aspects of campus life may benefit from the system; transponders prove equally effective in situations where a debit card for all in-campus services would simplify payment and bookkeeping procedures (e.g. Xerox, cafeteria, etc.). Depending on their storage capacity, transponders also allow the keeping of attendance and library records. Characterized by a high degree of generality, the system is easily customizable; depending on various needs, information from other campus areas may be selected to enter the system at a later time, e.g. book loan records, book codes, academic records, emergency medical advice and specifications.

## 4. Conclusions

RFID tags represent a reliable alternative to bar codes in product tracking and identification. Other advantages are worth mentioning here: an increased range of distance reading, speedy reading, sturdiness, high security and storage memory.

The paper presented an incorporated an RFID system designed to simultaneously control the access in different areas. The tags meet the requirements of the ISO 15693 standard for 13.56 MHz. Characterized by flexibility and modularity, the system may be installed with a minimum configuration and subsequently expanded to meet various needs. The graphic interface is user-friendly. The system eliminates human errors and ensures a high-quality management of inter-related aspects; it is very fast in processing data coming from thousands of transponders at thousands of access points.